\documentclass{aa}  

\usepackage{txfonts}
\usepackage{float}
\usepackage{graphicx}
\usepackage{xcolor}

\begin{document}

   \title{Localising pulsations in the hard X-ray and microwave emission of an X-class flare}

   \titlerunning{Localising QPPs in HXR and microwave flare emission}
    
   \subtitle{}

   \author{Hannah Collier \inst{1}\fnmsep\inst{2}\thanks{E-mail: hannah.collier@fhnw.ch}
          \and 
          Laura A. Hayes \inst{3}
          \and
          Sijie Yu \inst{4}
          \and
          Andrea F. Battaglia \inst{1}\fnmsep\inst{2}
          \and
          William Ashfield \inst{5,6}
          \and 
          Vanessa Polito \inst{6}
          \and
          Louise K. Harra \inst{2}\fnmsep\inst{7}
          \and
          S\"am Krucker \inst{1}\fnmsep\inst{8}
          }

        \institute{University of Applied Sciences and Arts Northwestern Switzerland (FHNW), Bahnhofstrasse 6, 5210 Windisch, Switzerland
        % \\
      % \email{hannah.collier@fhnw.ch} 
        \and
            ETH Z\"{u}rich,
              R\"{a}mistrasse 101, 8092 Z\"{u}rich Switzerland
        \and
            European Space Agency, ESTEC,
            Keplerlaan 1 - 2201 AZ, Noordwijk, The Netherlands
        \and
            New Jersey Institute of Technology, Newark, USA
        \and
            Bay Area Environmental Research Institute, NASA Research Park, Moffett Field, CA, 94035, USA
        \and 
            Lockheed Martin Solar \& Astrophysics Laboratory, 3251 Hanover Street, Palo Alto, CA, 94304, USA
        \and
             PMOD/WRC, Dorfstrasse 33, CH-7260 Davos Dorf, Switzerland
        \and
            Space Sciences Laboratory, University of California, 7 Gauss Way, 94720 Berkeley, USA
            }
        \authorrunning{Collier et al.}
        \date{Received November 17 2023; accepted February 15 2024}

% \abstract{}{}{}{}{} 
% 5 {} token are mandatory
 
    \abstract
  % context heading (optional)
  % {} leave it empty if necessary  
   {}
  % aims heading (mandatory)
   {This work aims to identify the mechanism driving pulsations in hard X-ray (HXR) and microwave emission during solar flares. Here, by using combined HXR and microwave observations from Solar Orbiter/STIX and EOVSA we investigate an X1.3 GOES class flare, 2022-03-30T17:21:00, which displays pulsations on timescales evolving from $\sim 7$ s in the impulsive phase to $\sim35$ s later in the flare.} 
  % methods heading (mandatory)
   {The temporal, spatial and spectral evolution of the HXR and microwave pulsations during the impulsive phase of the flare are analysed. Images are reconstructed for individual peaks in the impulsive phase and spectral fitting is performed at high cadence throughout the first phase of pulsations.} 
  % results heading (mandatory)
   {Imaging analysis demonstrates that the HXR and microwave emission originates from multiple sites along the flare ribbons. The brightest sources and the location of the emission changes in time. Through HXR spectral analysis, the electron spectral index is found to be anti-correlated with the HXR flux showing a ``soft-hard-soft'' spectral index evolution for each pulsation. The timing of the associated filament eruption coincides with the early impulsive phase.}
  % conclusions heading (optional), leave it empty if necessary 
   {Our results indicate that periodic acceleration and/or injection of electrons from multiple sites along the flare arcade is responsible for the pulsations observed in HXR and microwave. The evolution of pulsation timescales is likely a result of changes in the 3D magnetic field configuration in time related to the associated filament eruption. } 


\maketitle
%
%-------------------------------------------------------------------

\section{Introduction}
Rapid variations, on the order of seconds to tens of seconds, are often observed in the amplitude of emission from solar flares (sometimes classified as Quasi-Periodic Pulsations or QPPs \citep{zimovets2021}). These variations are present in all wavelengths of emission. In particular, they are often present clearly in the non-thermal Hard X-ray (HXR) emission during a flare. The brightest HXR emission observed is a result of the interaction of non-thermal flare-accelerated electrons with chromospheric plasma to produce non-thermal bremsstrahlung emission. In this sense, HXR observations enable the study of particle acceleration and transport in solar flares. Microwave observations are a complementary dataset to HXR because at microwave wavelengths the population of electrons which is trapped in the coronal magnetic flux tube is probed and this is useful when considering acceleration and transport effects. QPPs and time-variations are often present in flare emission at microwave wavelengths and show similar signatures to HXR. QPPs have been identified with timescales ranging from seconds to minutes \citep[e.g.][]{2023SCPMA..6659611Z, 2022NatCo..13.7680K, 2022FrASS...940945L, 2022FrASS...932099L, 2022Ge&Ae..62..356Z}, with few studies reporting sub-second QPPs \cite[e.g.][]{Knuth_2020, QIU_2012}. The most commonly reported pulsation timescales are on the order of seconds - tens of seconds \cite[e.g.][]{Inglis_2016, Hayes_2020}, which are particularly relevant for particle acceleration studies. In many cases the pulsation timescales are shown to evolve over the course of the flare \citep{hayes2019, dennis2017}. A recent statistical study of QPPs by \cite{2023MNRAS.523.3689M} found that 81\% of flares displaying QPPs in both the impulsive and decay phase showed non-stationarity, meaning that the characteristic timescales of flare pulsations evolved in time during the course of the flare. 

To explain the underpinning mechanism driving solar flare pulsations, various models have been proposed. For recent reviews of the proposed models we refer the reader to \citet{zimovets2021, Kupriyanova2020quasi-periodic,mclaughlin2018, vandoorsselaere_2016, nakariakov2009}. Typically these models involve direct modulation of the plasma due to magnetohydrodynamic (MHD) oscillations in a flaring loop, periodic energy release driven by MHD modes, or energy release process that have an intrinsic characteristic timescale or period. From an observational perspective, it is challenging to distinguish between possible drivers.  A review article by \cite{zimovets2021} emphasises the importance of spatially resolving QPP sources and studying their dynamics at different energy ranges.

Previous works have attempted to localise the source of pulsations \cite[e.g.][]{2003ApJ...588.1163G, 2005A&A...439..727M, Clarke_2021, 2022NatCo..13.7680K}. Specifically, \cite{Clarke_2021} studied an M3.7 GOES class flare displaying pulsations with periodicities of $\sim137^{+49}_{-56}$ s in the HXR emission. In this work the location of QPPs was found to be along the flare ribbons. The periodic signal from a HXR footpoint close to a system of open field lines was particularly apparent. As a result, periodic type III radio bursts were also observed. This was determined primarily through spatial analysis at extreme ultraviolet (EUV) wavelengths observed by the Atmospheric Imaging Assembly (AIA) onboard the Solar Dynamics Observatory (SDO). \cite{Fleishman_2008} studied radio and X-ray pulsations in an X class flare. Through detailed analysis which involved studying the degree of polarisation and spectral index variation, among other features, the authors concluded that periodic injection and/or acceleration of electrons was the most likely cause of X-ray and radio pulsations. In that work spatial resolution was a limiting factor. Here, we take advantage of the new HXR observations from the Spectrometer Telescope for Imaging X-rays (STIX) on-board Solar Orbiter \citep{2020A&A...642A...1M}, together with microwave observations from the Expanded Owens Valley Solar Array (EOVSA) to study the temporal, spectral, and spatial properties of QPPs observed in an X-class solar flare. 

STIX is a HXR imaging spectrometer with a 1 keV resolution (at 6 keV), and detects photons with energies in the range 4-150 keV \citep{2020A&A...642A..15K}. Importantly, STIX has a high time resolution of 0.5 s and continuously observes the full solar disk from a unique vantage point offered by the trajectory of Solar Orbiter. The spacecraft reaches a distance of 0.3 AU from the Sun at perihelion. These capabilities mean that STIX is a suitable instrument for analysing rapid variation in the HXR emission from flares. It is important to note that STIX is an indirect Fourier imager, similar to the Reuven Ramaty High Energy Solar Spectroscopic Imager (RHESSI) \citep{Lin2002}, which has certain limitations that will be discussed in the sections to follow. EOVSA is a ground-based radio telescope array that provides high time cadence observations at 1 s \citep{Gary_2018} and samples microwave emission in the frequency range 1-18 GHz. EOVSA is similarly well suited for the study of temporal variations on second timescales in flare emission. These two instruments with the support of EUV/UV observations from AIA \citep{2012SoPh..275...17L} are used in this work to localise the source of QPPs in an X1.3 GOES class flare which was observed during the March 2022 perihelion of Solar Orbiter. 

Section \ref{sec:observations} details the observations obtained for the flare studied in this work. In section \ref{sec:results}, the key results obtained are presented, including results from imaging and spectral analysis. Section \ref{sec:discussion} discusses the results and their limitations, as well as potential QPP models which could explain the observations. Finally, section \ref{sec:conclusions} gives the conclusions derived from this work. 

%--------------------------------------------------------------------
\section{Observations}\label{sec:observations}
\begin{figure*}
    \centering
    \includegraphics[width=\textwidth]{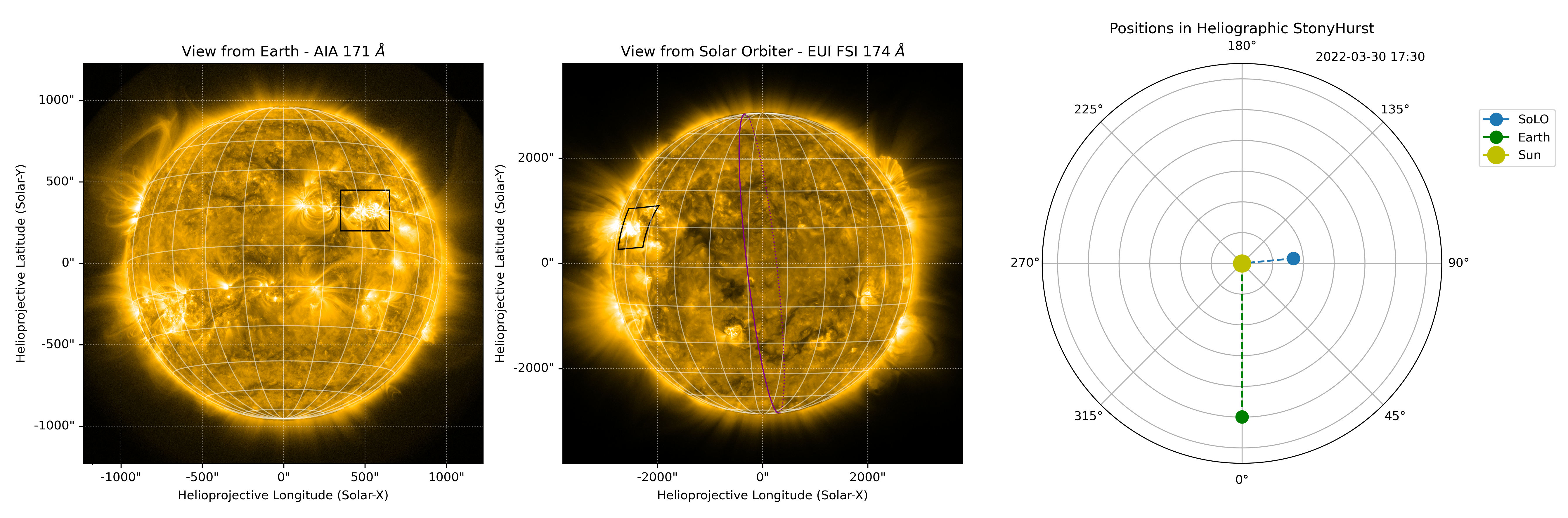}
    \caption{An overview of the field of view from both Earth and Solar Orbiter and the location of the Solar Orbiter spacecraft with respect to Earth on March 30th 2022. The leftmost panel shows an AIA 171~\AA\ image at 2022-03-30 17:29:57. The flaring active region of interest lies within the black box. The middle panel shows a 174~\AA\ image from the Full Sun Imager (FSI) on-board Solar Orbiter's Extreme Ultraviolet Imager (EUI) \citep{EUI_instrument_paper} from 17:36:20. The limb as seen from AIA is shown in purple. The rightmost panel shows a top-down view of the spacecraft location with respect to the Sun-Earth line in Heliographic Stonyhurst coordinates.}
    \label{fig:spacecraft_location}
\end{figure*}

In this work the X1.3 GOES class flare which occurred on March 30th 2022 is studied. At this time Solar Orbiter was near its perihelion and was at a distance of 0.33 AU from the Sun with an angular separation of $95^{\circ}$ to the Sun-Earth line as shown in Fig 
\ref{fig:spacecraft_location}. The flare occurred towards the East limb as observed from Solar Orbiter and towards the West limb from the reference frame of Earth. The black box in Fig \ref{fig:spacecraft_location} denotes the flaring active region from both viewpoints.

\begin{figure*}
     \centering
    \includegraphics[width=\textwidth]{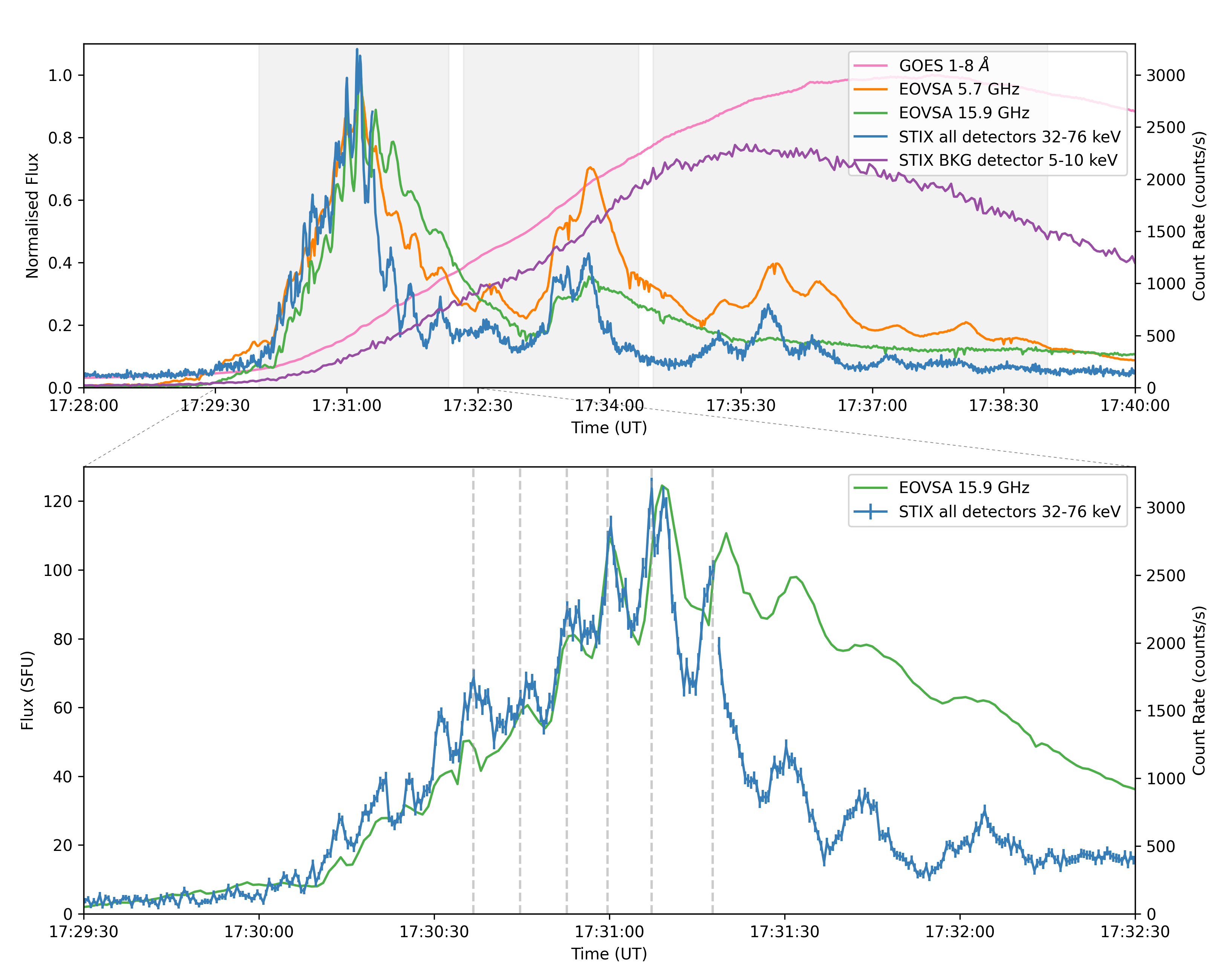}
    \caption{Overview plot of the event showing the time profiles from several instruments. The top panel shows the non-thermal and thermal evolution of the flare from STIX, EOVSA and GOES/X-ray Sensor (XRS). The top panel shows the normalised GOES/XRS lightcurve alongside the normalised microwave emission observed with EOVSA in the 5.7 and 15.9 GHz channels. The STIX 32-76 keV time profile is shown, where the flux is summed over all detectors. The STIX 5-10 keV lightcurve is shown as observed by the background detector (BKG). The bottom panel shows a zoom in of the early impulsive phase of pulsations (phase 1) seen in HXR and microwave emission. The times shown are given at Earth in UT. The dashed lines correspond to the time at the centre of the integration bin used for reconstruction of the HXR images shown in Fig. \ref{fig:early_impulsive_Earth}. They correspond to the mean times of each component derived from the Gaussian decomposition method presented in \cite{collier_2023}.}
    \label{fig:overview_figure}
\end{figure*}

\begin{figure*}
    \centering
    \includegraphics[width=\textwidth]{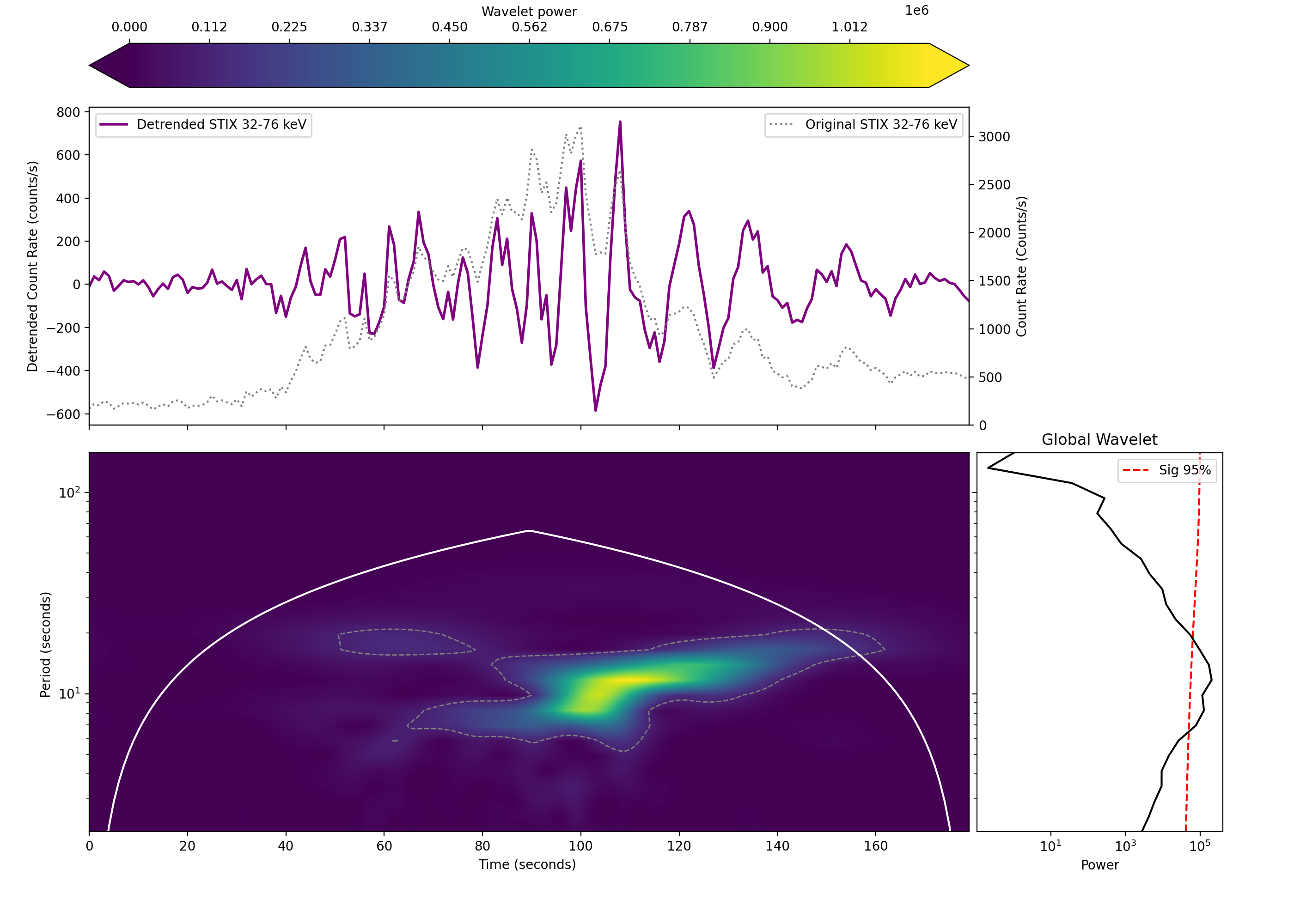}
    \caption{Wavelet power spectrum for the detrended lightcurve of the impulsive phase of HXR emission. The time range used is the same as in Fig. \ref{fig:overview_figure}. The dotted grey curve in the top panel is the original lightcurve prior to detrending. The bottom panel shows the wavelet power spectrum which has enhanced power above the 95\% significance level (the dashed white curve) during the early impulsive phase. The global wavelet is also shown on the right hand side. The enhanced power is at a period of $\sim 10$ s, which agrees well with the 7 s period derived from the Gaussian decomposition method in \cite{collier_2023}.}
    \label{fig:wavelet}
\end{figure*}

The flare exhibits interesting pulsations in the HXR emission observed by STIX throughout the impulsive phase and past the peak of thermal emission (see Fig. \ref{fig:overview_figure}). The ground software used for STIX imaging and spectral analysis is version 0.4.0\footnote{https://github.com/i4Ds/STIX-GSW}. The HXR time profile presented in Fig. \ref{fig:overview_figure} has been live time corrected using the current best correction factors for total dead time (ASIC $\tau =  1.1 
~\mu{s}$ and FPGA $\tau = 10.1 ~\mu{s}$). It is also important to note that all times given in this work are in UTC at Earth. There are three main phases of pulsations (the three shaded regions in Fig. \ref{fig:overview_figure}) which were characterised in previous work by \cite{collier_2023}. These were determined by decomposing the signal into individual Gaussian bursts. The three phases display quickly varying behaviour on timescales growing from just $\sim 7$ s in the early impulsive phase to $\sim 35$ s in the third phase. In this work we particularly focus on the first phase of pulsations because they behave remarkably periodically as demonstrated by the wavelet transform shown in Fig. \ref{fig:wavelet} for which a Morlet wavelet was used\footnote{Wavelet software was provided by C. Torrence and G. Compo and is available at: http://atoc.colorado.edu/research/wavelets/. }. 

As well as observations from STIX, the flare was observed by Earth based instruments including the ground based radio interferometer, EOVSA. The time profiles of the microwave observations from EOVSA are also presented in Fig.~\ref{fig:overview_figure}, for two select frequencies, 5.7 and 15.9~GHz. The EOVSA observations are of particular interest as the emission observed is predominantly generated by accelerated electrons in the corona via gyrosynchrotron emission. Various studies have shown that gyrosynchrotron microwave and HXR bremsstrahlung emission can originate from the same population of flare accelerated electrons and the observed emission is often well correlated \cite[e.g.][]{Krucker_2020_micro, 1990A&A...229..206A}. Thus, microwave observations are a complementary probe of accelerated electrons in a flare \citep{white2011}. In this case, the microwave emission observed by EOVSA shows pulsations which correlate remarkably well with those seen in HXR. This is highlighted in the bottom panel of Fig~\ref{fig:overview_figure}. However, after the non-thermal peak (17:31:10), each peak in microwave becomes less symmetric, unlike the the HXRs peaks. Interestingly, the final marked HXR peak in the bottom panel of Fig~\ref{fig:overview_figure} is clearly out of phase with the microwave peak. The HXR peak notably reaches its maximum ahead of the microwave peak. This is particularly prominent in the 15.9~GHz profile, but is also the case at lower frequencies. This is typically attributed to electron trapping \citep{2000ApJ...545.1116S, 2001ApJ...547.1090K}. It is important to note that the STIX attenuator was inserted at 17:31:18. This complicates the HXR live time correction and therefore the data point at this time is not shown. However, in depth analysis of this aspect is beyond the scope of this work.

In this work, we expand upon the analysis performed in \cite{collier_2023} and study the spatial and spectral evolution of the pulsations. We first begin by reconstructing the HXR and microwave images of the pulsations and then analyse the spectral evolution on these timescales. Finally we relate the observations back to a more general picture of the filament eruption which occurred (see the movie in additional material). These observations were used to disambiguate between potential QPP mechanisms and as a result we identify periodic injection and/or acceleration of electrons as the driver of time variation in the observed HXR and microwave emission. 

\section{Results} \label{sec:results}

\subsection{Hard X-ray \& Microwave Imaging}
The reconstruction of reliable images using an indirect Fourier imager requires significant flux for signal modulation. One way to increase the signal to noise ratio is by increasing the image integration time. A compromise must be found so that one can distinguish between the HXR emission from an individual burst whilst having sufficient flux to produce a reliable image. For details on the STIX imaging concept we refer the reader to \cite{2023SoPh..298..114M}. For this flare, HXR images were reconstructed for each of the Gaussian pulses fitted in \cite{collier_2023}. The integration times used were the Full Width at Half Maximum (FWHM), centered around the mean of the fitted Gaussian pulses. The time intervals are given in Table \ref{tab:integration_times}. The integration times are relatively short, however, they were deemed sufficient given the large flux of this flare. There are between 23,098 and 31,637 counts in each image (see Table \ref{tab:integration_times}). For comparison, \cite{2023A&A...670A..89S} reliably reconstructed a 22-28 keV STIX image of four HXR sources of similar intensities with 11,189 counts. Here, there are excellent counting statistics, with double and in some intervals, nearly triple the counts. 

\begin{table*}
    \centering
    \begin{tabular}{|c|c|c|c|}
    \hline
         Peak No. & Time Range (UT at Earth) & Integration Time (s) & Counts above Background (20-76 keV)\\
         \hline
         1 & 17:30:32.2 - 17:30:41.2 & 9 & 23098 \\
         \hline
         2 & 17:30:39.2 - 17:30:50.2 & 11 & 28887 \\
         \hline
         3 & 17:30:48.2 - 17:30:57.2 & 9 & 25638 \\
         \hline
         4 & 17:30:55.2 - 17:31:04.2 & 9 & 27097 \\
         \hline
         5 & 17:31:02.2 - 17:31:12.2 & 10 & 29961 \\
         \hline
         6 & 17:31:11.2 - 17:31:24.2 & 13 & 31637 \\
         \hline
    \end{tabular}
    \caption{Integration times used for the reconstruction of HXR images shown in Fig. \ref{fig:early_impulsive_Earth}. They correspond to the FWHM time ranges for the Gaussian components fit in \cite{collier_2023}. The total counts above background in the 20-76 keV energy range for each interval is given.}
    \label{tab:integration_times}
\end{table*} 

Due to the relative position of this flare to the STIX grids (this flare was at the East limb from Solar Orbiter's vantage point), the Caliste-SO detectors which consist of 12 pixels (four top, four bottom and four small pixels), as described in \cite{2020A&A...642A..15K}, were not fully illuminated. In particular, the top pixels were partially covered and measured approximately 85\% of the flux of the bottom detectors. This effects the resultant Moiré pattern formed on the top pixels. We therefore only use bottom pixels for image reconstruction. This reduces the counts available for reconstruction.

\begin{figure*}
     \centering
    \includegraphics[width=\textwidth]{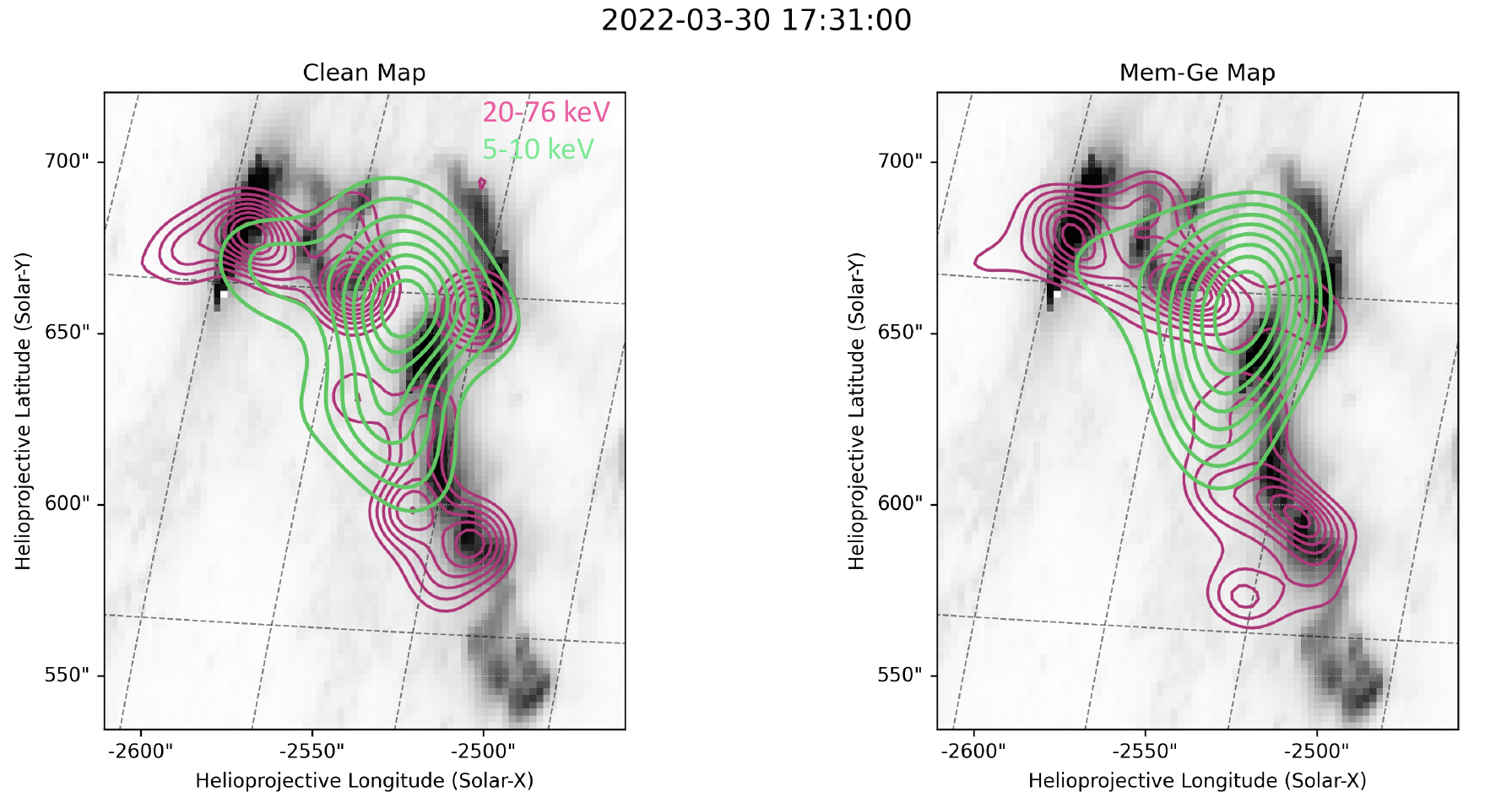}
    \caption{HXR Clean and MEM\_GE images for the peak centred at 17:31:00 overlaid on the AIA 1600~\AA\ map at 17:31:02, in the Solar Orbiter reference frame. The pink contours at 20-90\% represent the 20-76 keV STIX maps and the green 20-90\% contours represent the 5-10 keV STIX maps. Both maps show HXR emission along the UV flare ribbons. The Clean algorithm produces more fractured HXR bright points compared to the smeared out emission obtained with MEM\_GE. These multiple HXR sources correspond remarkably well to bright points observed along the UV ribbons.}
    \label{fig:early_impulsive_solo}
\end{figure*}

Fig. \ref{fig:early_impulsive_solo} shows background subtracted reconstructed HXR images for the peak centred at 17:31:00 (peak 4 in Table \ref{tab:integration_times}), for both the 5-10~keV and 20-76~keV energy bands in green and pink, respectively. The underlying AIA 1600~\AA\ map is the frame closest in time to the centre of the interval of the STIX image and was taken at 17:31:02. The figure is shown in the Solar Orbiter reference frame; the AIA map was reprojected to this frame using the reproject functionality provided by SunPy \citep{sunpy_community2020}. For the non-thermal images subcollimators 3-10 were used, which correspond to subcollimator resolutions of 14-178'', because there was little-to-no modulation in the finest resolution subcollimators. Correspondingly, only subcollimators 5-10 were used to reconstruct the thermal maps. The figure on the left of Fig. \ref{fig:early_impulsive_solo} shows STIX maps reconstructed by the Clean algorithm \citep{hurford2002SoPh..210...61H} and on the right by the MEM\_GE algorithm \citep{Massa_2020}. Detail on the fit of the reconstructed non-thermal maps to the observed visibilities is given in appendix \ref{apdx:vis_fits}. Finally, we note that a shift of (-13, 45)'' has been applied to the STIX maps by manually aligning the emission to that from AIA 1600 {\AA}, when reprojected to the Solar Orbiter viewpoint. This shift is necessary due to the currently achieved accuracy of the STIX aspect system. The same shift is used in STIX maps shown throughout this paper.  

It is clear from Fig. \ref{fig:early_impulsive_solo} that at this time the non-thermal HXR emission originated from locations all along the ultraviolet (UV) flare ribbons. The main difference between the Clean and MEM\_GE reconstructions is that for Clean the distribution of bright points is fractured along the ribbons, whereas MEM\_GE gives a smoother, smeared out distribution of emission across the flare ribbons. The algorithm results in this effect since the Clean components are convolved with a narrow beam. Therefore, it is important to ensure an appropriate beam size choice is made, so as to not over-resolve/separate bright points. Here, a clean beam size of 16.5'' was used for the non-thermal map which is slightly larger than the resolution of the finest subcollimator used in image reconstruction i.e. a conservative approach was taken when choosing the Clean beam width. This results in localisation of HXR bright points which correspond remarkably well to the fragmentation of UV brightenings observed along the ribbons in AIA 1600~\AA\ . 

\begin{figure*}
        \centering
    \includegraphics[width=\textwidth]{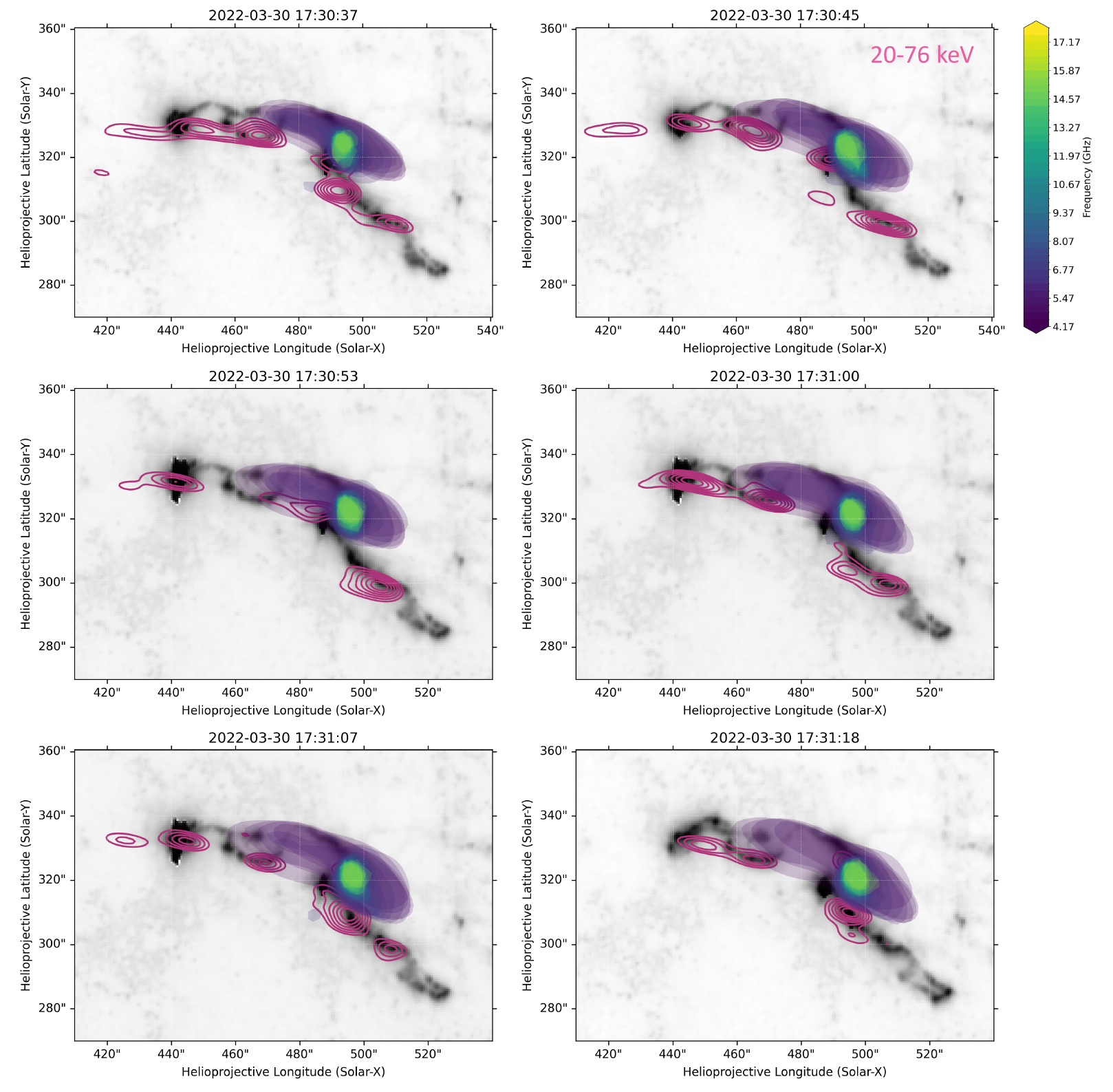}
    \caption{AIA 1600~\AA\ maps with 20-76 keV STIX Clean map 40-90\% contours overlaid and 60 - 90\% EOVSA microwave contours observed at frequencies ranging from 4-18 GHz, with the colour map ranging from purple to yellow in increasing frequency. Each image corresponds to a single HXR peak in phase 1 as determined by the Gaussian decomposition method in \cite{collier_2023}. The integration time for each image is the FWHM of the fitted Gaussian burst. The time shown above each frame is the mean time of the burst at Earth in UTC which are indicated by vertical dashed lines in Fig. \ref{fig:overview_figure}. The AIA maps closest in time after the centre of each STIX image interval is shown.}
    \label{fig:early_impulsive_Earth}
\end{figure*}

To compare microwave and HXR observations, we need to analyse the two datasets in the same reference frame. Since HXR footpoint emission is known to originate from the chromosphere and the altitude of microwave emission is not well constrained, we reproject STIX observations to Earth. Figure~\ref{fig:early_impulsive_Earth} shows the Clean images for individual peaks in the early impulsive phase of pulsations. Each frame corresponds to the time intervals specified in Table \ref{tab:integration_times}, the centre of which is denoted by the dashed lines in Fig. \ref{fig:overview_figure}. The STIX maps are shifted as described previously and then reprojected to Earth's coordinate frame. It is not sensible to reproject the soft X-ray sources (5-10~keV) because their altitude in the corona is also poorly constrained, therefore we only show the non-thermal emission in this case. The 20-76~keV HXR 40-100\% contours during the impulsive phase are shown in pink overlaid on the AIA 1600 {\AA} maps that are closest in time to the centre of the integration time used for STIX maps. Here 40\% is the lowest contour level displayed because the first few frames have fewer counts than the interval presented in Fig. \ref{fig:early_impulsive_solo} and thus a lower signal-to-noise ratio.

EOVSA provides microwave images ranging from 1--18 GHz, employing 451 science channels distributed across 50 spectral windows. The flux, bandpass and complex gain calibrations are executed through the standard EOVSA imaging pipeline. Following this standard calibration, an extra self-calibration round is initiated to address any residual phase/amplitude discrepancies in the calibrated data. Such discrepancies can arise from factors such as atmospheric density variations, changes in antenna-based gain, among others \citep{Cornwell1999}. For microwave observations, we subtract the pre-flare background visibilities, averaged over a 20-second interval starting at 17:28:40 UT, from the observed visibilities. In this study, microwave imaging is carried out on the background-subtracted visibility data for every spectral window within the 3.5 to 18 GHz range. This results in images at 45 uniformly spaced frequencies with a 2-second time cadence. The images were then reconstructed using a circular beam, characterized by a full-width-half-maximum (FWHM) size of $60''/\nu_{\rm GHz}$, where $\nu_{\rm GHz}$ represents the image frequency in GHz. The image intensity is adjusted by calibrating the integrated flux across the image plane with the total power flux derived from a single-dish measurement. The calibration of image-based flux is performed separately for each spectral window. Fig. \ref{fig:early_impulsive_Earth} shows the 60 - 90\% EOVSA microwave contours observed at frequencies ranging from 4-18 GHz, with the colour map ranging from purple to yellow in increasing frequency. High contour levels were chosen for display here so as to not mask the HXR contours. During phase 1, microwave images at frequencies $\gtrsim 10$ GHz (denoted by the green to yellow contours) display a compact source near the center of the flare ribbons. At lower frequencies (spanning from purple to green), the microwave source exhibits an elongated shape tracing the UV flare ribbons. 

Fig. \ref{fig:early_impulsive_Earth} demonstrates that HXR emission is present at multiple locations across the UV flare ribbons during the first phase of pulsations. The location of the brightest emission changes in time and many sources are present in each frame. The precise location of HXR emission evolves in time. The source appearing at the most eastward location is exaggerated due to projection effects and doesn't appear to have a corresponding UV brightening. 

We attempted to forward fit the HXR visibilities for the time range shown in Fig. \ref{fig:early_impulsive_solo}, however, fitting five circular Gaussian sources involves too many free parameters (20) to be accurately fit with 24 visibilities, especially in the case where some subcollimators do not resolve the source sizes. This issue needs to be studied further in future work and is beyond the scope of this work. In any case, the non-thermal source structures in this flare are very complex with multiple sources present at the same time. STIX is not designed to be able to fully reconstruct such complex source geometries. As a result, the reconstructed images only show the most prominent sources, while fainter sources are lost in the limited dynamic range of the reconstructed images. 

\begin{figure}
    %\centering
    \includegraphics[width=\columnwidth]{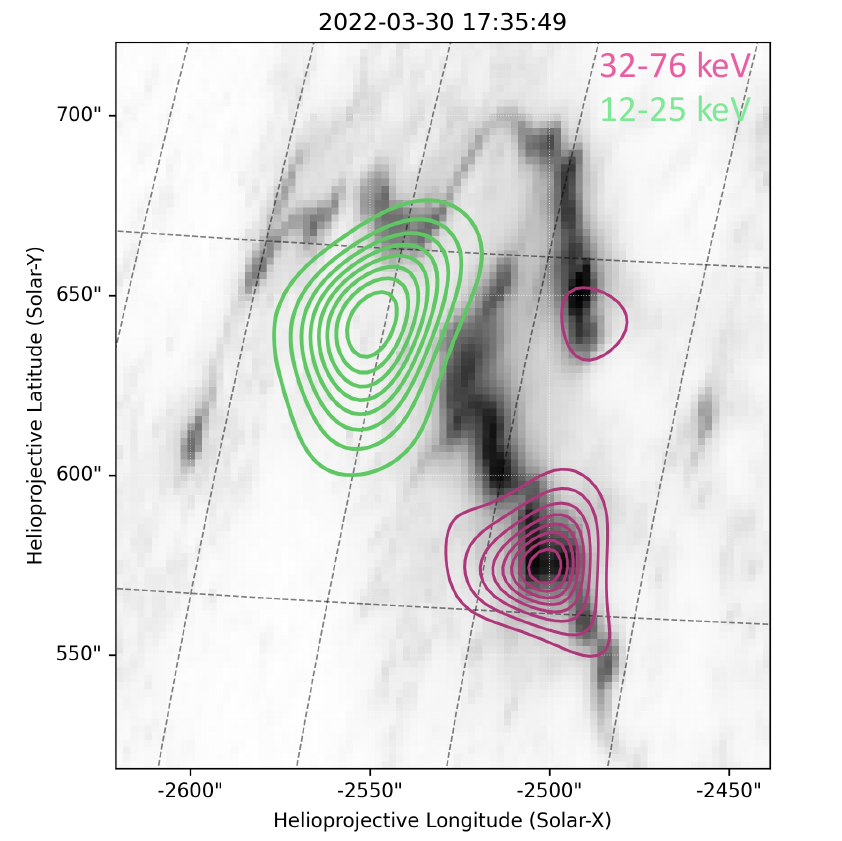}
    \caption{An AIA 1600~\AA\ map with 12-25 keV thermal and 32-76 keV non-thermal MEM\_GE STIX contours overlaid for the third phase of pulsations. The image is shown in the Solar Orbiter reference frame. The centre of the image interval from STIX is at 17:35:49. At this late phase the standard flare cartoon with two non-thermal footpoints and connecting thermal loop-top emission represents the observations well.}
    \label{fig:phase3_imaging}
\end{figure}

Fig. \ref{fig:phase3_imaging} shows the HXR observations at a later time (17:35:49) during the third phase of pulsations. The energy ranges used are slightly different to previous intervals since at this time the relative contribution of thermal to non-thermal emission is higher. As such the non-thermal images are from 32-76 keV and the thermal maps from 12-25 keV. In addition, for the non-thermal and thermal images subcollimators 4-10 and 5-10 were used, respectively. The HXR sources are over-plotted at the 20-90\% contours levels on a AIA 1600 {\AA} image closest in time to the centre of the STIX image. It is notable that the HXR sources are now located further towards the South-West along the ribbon. In contrast to earlier times, the standard flare picture better represents reality, with only two footpoints and a connecting loop structure present.

\subsection{Spectral Analysis}\label{subsec:spectral_analysis}

\begin{figure*}
    \centering
    \includegraphics[width=\textwidth]{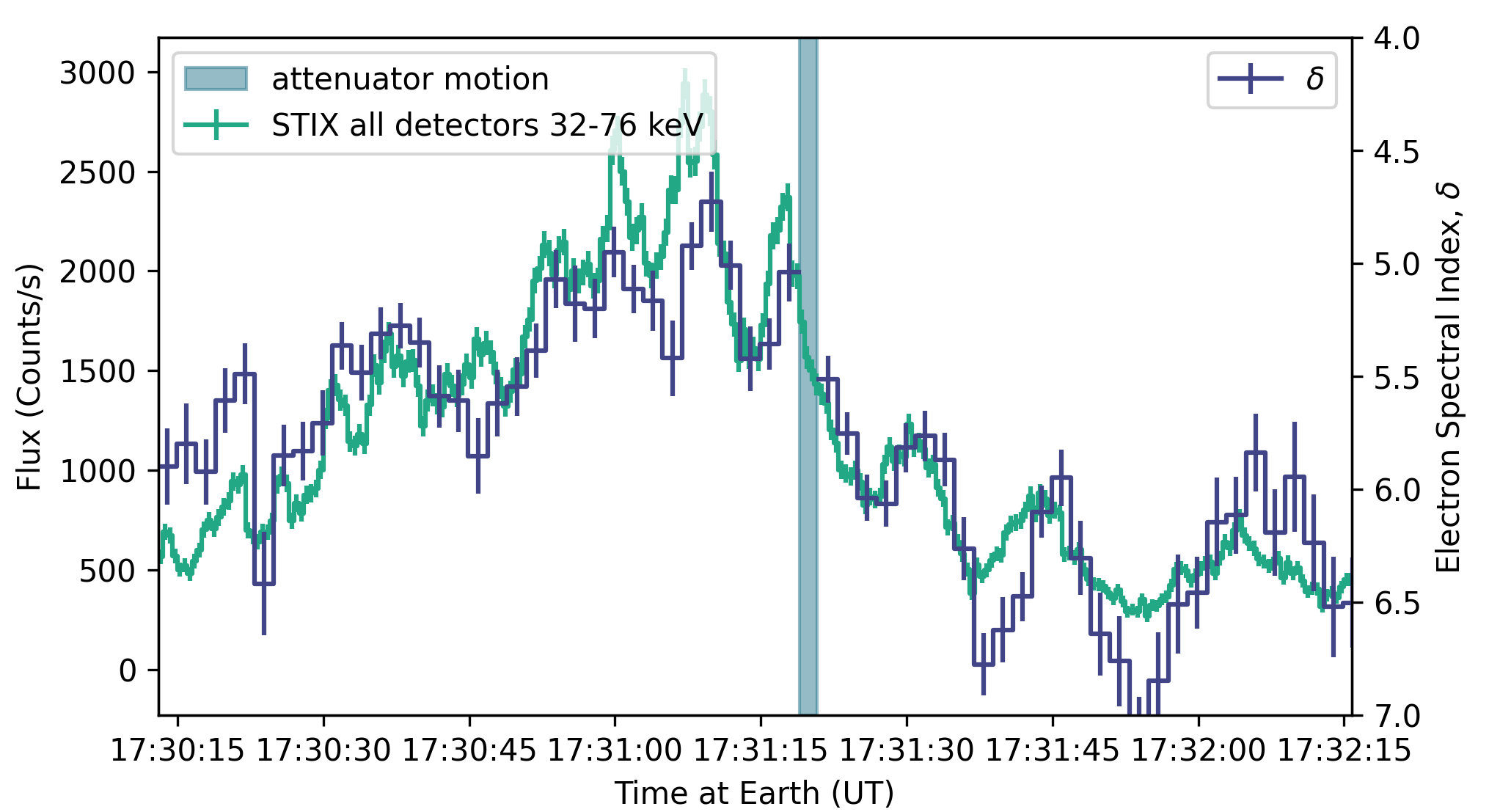}
    \caption{32-76 keV HXR flux observed by STIX during the first early impulsive phase. The electron spectral index evolution in time is also shown. The electron spectral index is anti-correlated with the observed flux and shows a soft-hard-soft evolution with each pulse. This indicates that the pulsations are related to the acceleration of electrons.}
    \label{fig:phase_1_spectral_evolution}
\end{figure*}

We analyse the X-ray spectra measured by STIX to gain insights into the spectral evolution of the observed pulsations. Here, we fit the HXR spectra using \texttt{OSPEX} with both a thermal (\texttt{f\_vth.pro}) and thick-target bremsstrahlung (\texttt{f\_thick2.pro}) component over the energy range of 10-63~keV at 2 s intervals. The fits did not include counts below 10 keV because during the impulsive phase the attenuator was inserted. In order to simplify the fitting procedure and for consistency we fit above 10 keV as the instrumental response in this energy range is well understood. Since only the derived spectral index is required for our analysis, the accuracy of the thermal fit is not pertinent to this study. An example fit is shown in appendix \ref{apdx:example_spectral_fit}. Only counts falling on the bottom pixels of each detector were used because of the shadowing effect discussed previously. From the non-thermal thick-target fit  the electron spectral parameters were derived. In particular, the electron spectral index as a function of time was obtained. Fig.~\ref{fig:phase_1_spectral_evolution} shows the electron spectral index time evolution as determined by STIX for the first phase of pulsations compared to the observed HXR flux. We note that the electron spectral index is anti-correlated with the HXR flux and obeys a soft-hard-soft relationship with each pulsation. This indicates that each pulsation is associated with a new acceleration and/or injection of an electron population into the flaring loops. 
This suggests that the mechanism for the observed rapidly varying behaviour must be able to modulate the electron spectral index significantly, either by supplying significant amounts of energy to electrons in an energy dependent way or by injecting a new population of electrons. Fig. \ref{fig:phase_1_spectral_evolution} shows this relationship for phase 1, when there are pulsations on short timescales ($\sim 7$ s), however, this soft-hard-soft relationship also continues into the later phases (2 \& 3), with the baseline spectral index gradually becoming harder \citep{Grayson_2009} i.e. the global trend shows a soft-hard-hard behaviour. 

\begin{figure*}
    \centering
    \includegraphics[width=\textwidth]{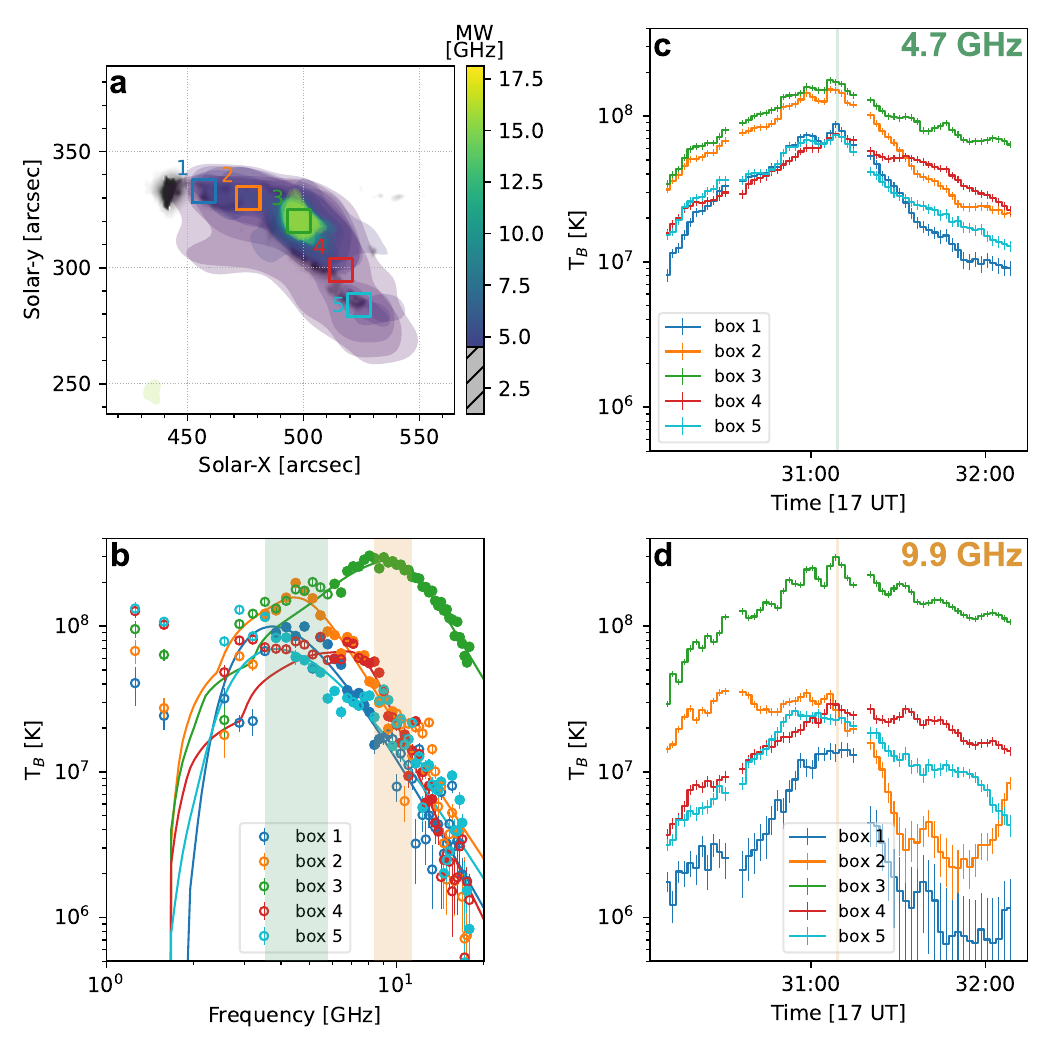}
    \caption{Spatially-resolved microwave spectral analysis for Phase 1. Panel (a) displays the AIA 1600 \AA\ maps overlaid with 25\% contours from the EOVSA microwave map, ranging from 3.5-18 GHz, captured at 17:31:08 UT (corresponding to peak No. 5 in Table \ref{tab:integration_times}). The five distinct boxes mark regions selected for spectral fitting. Panel (b) illustrates the brightness temperature spectra from the five regions at 17:31:08 UT. Spectra are color-coded to match the boxes in Panel (a). The color-matching curves (in green and orange colors) represent best-fit models derived from homogeneous gyrosynchrotron emissions due to non-thermal electrons with a single power-law distribution. The open circles denotes data points excluded for the spectral fit. Panels (c) and (d) provide EOVSA spatially-resolved time profiles from the five chosen regions, each averaged over specific frequency ranges. Panel (c) emphasizes the range 3.5--5.8 GHz, as highlighted by the green shaded area in Panel (a), with a median frequency of 4.7 GHz. Similarly, Panel (d) centers on the frequency band between 8.4--11.3 GHz, marked by the yellow shading in Panel (a), with a median frequency of 9.9 GHz.}
    \label{fig:microwave_fitting} 
\end{figure*}

In conjunction with the HXR spectral analysis, we have derived the spatially-resolved brightness temperature spectra from five selected locations along the microwave source bridging the two ends of the UV flare ribbons. The associated error bars were calculated by combining the root mean square noise level in the image with an assumed systematic error of 10\% of the absolute brightness temperature, computed in quadrature. At the peak of phase 1 (peak 5 in Table \ref{tab:integration_times}), the spatially resolved microwave spectra from all five sources, as illustrated in Fig. \ref{fig:microwave_fitting}, exhibit characteristics of nonthermal gyrosynchrotron radiation \citep{dulk1985ARA&A..23..169D}. The closed circles in Fig. \ref{fig:microwave_fitting} denote data points included in fitting. The spectra generally reveal a negative and/or positive slope at the high- and low-frequency sides (attributed to the optically thin and thick non-thermal gyrosynchrotron emissions, respectively), with a peak brightness temperature exceeding 300 MK at the center and $\lesssim$ 100 MK at other regions. We note that the spectra corresponding to the central region (green) and the adjacent region to the right (red) display secondary spectral peaks, suggesting an inhomogeneous emission source, whose spectral shape is not consistent with a homogeneous source model. This inhomogeneity could stem from various factors, including the existence of a secondary electron population or fluctuations in magnetic field strength in these regions. Consequently, we have excluded the data points at frequencies that correspond to the secondary peak for spectral fitting. 

It is important to note that gyrosynschrotron microwave emissivity depends on many factors. The relationship between gyrosynchrotron microwave emissivity $\eta$ and various factors including magnetic field strength $B$, angle $\theta$ to the magnetic field, and power law electron spectrum slope $\delta$, is approximated by \citet{dulk1985ARA&A..23..169D} as shown in Equation \ref{eq:dulk}.
\begin{equation}
    \eta(\nu, \theta, \delta) \propto 10^{-0.52\delta}B~n_e~(sin\theta)^{-0.43+0.65\delta}\left(\frac{\nu}{\nu_{B}}\right)^{1.22-0.9\delta}
    \label{eq:dulk}
\end{equation}
The brightness temperature spectra distinctly reveal that the central flare arcade exhibits the highest turnover frequencies relative to other regions, suggesting higher magnetic field strength in the region. Employing the fast gyrosynchrotron codes from \citet{2010ApJ...721.1127F}, we calculated the gyrosynchrotron brightness temperature spectrum from a homogeneous source, involving various model parameters such as magnetic field strength $B$ and power law index $\delta$ for the non-thermal electron distribution. A forward-fitting approach, detailed in \citet{2020Sci...367..278F}, was applied to reconcile the calculated model spectra with the observed ones. This fitting procedure was conducted separately for the spectra acquired at the five regions. Our findings indicate magnetic field strengths of approximately 850 G in the vicinity of the central flare arcade, which decreases to about 100 G in the outer regions of the flare ribbons.  These results are consistent with the imaging results that the majority of high-frequency microwave emission comes from the central region owing to the field strength dependence of emissivity shown in equation \ref{eq:dulk}. The distribution of magnetic field strength is also consistent with observations of photospheric magnetic fields along the ribbons, as detected by SDO's Helioseismic and Magnetic Imager (HMI). 

The non-thermal electron spectral index $\delta$, primarily constrained by the optically thin linear regime at frequencies above the turnover, demonstrates a harder slope in the central region. It ranges between 2.5 and 3, compared to a softer slope of approximately 4 in the outer regions. A large effort was made to derive the electron spectral index evolution in time from microwave observations. However, the accuracy of the power-law index $\delta$ relies heavily on the data collected at frequencies beyond the turnover, particularly within the linear regime of the power-law slope. For phase 1, the data is predominantly influenced by emissions from the central arcade, characterized by a strong magnetic field. Consequently, the dataset within the optically thin linear regime is somewhat limited and subject to variation over time. This variability directly affects the estimated power-law index values and their associated uncertainties, which are considerable and yield inconclusive results. Hence, these findings are not included here.

In panels (c) and (d) of Fig. \ref{fig:microwave_fitting}, the spatially resolved light curves at 4.7 and 9.9 GHz, derived from the five marked regions, are presented. Each data point on the light curves corresponds to the peak brightness temperature within its respective region. Despite the observed variations in brightness temperature levels, a synchronous temporal evolution at 4.7 GHz is evident from the onset of phase 1. This synchronised activity peaks collectively at the instance of peak No. 5 in Table \ref{tab:integration_times} and is followed by a clear decay before transitioning into phase 2. 
Pulsations on a $10$ s timescale appear to be present, however they are not particularly prominent when considering the systematic uncertainty at 4.7 GHz. Nonetheless, the coherence in the temporal and spectral characteristics across these spatially separated locations suggests potential magnetic connectivity. 

The 9.9 GHz light curves demonstrate clearer temporal behaviours. Distinct pulsations are observed in the central flare arcade, yet the situation for the adjacent regions is less obvious. In the left region (colored orange), some temporal variation on short timescales is apparent before the peak time which later diminishes, while in the right region (colored red), pulsations appear more prominently after the peak. The peripheral regions with weaker magnetic field strengths and lower emission intensities at high frequencies do not exhibit clear $10$ s pulsation patterns at 9.9 GHz.

\section{Discussion}\label{sec:discussion}
\subsection{Interpretation of Results}
In this flare three phases of HXR and microwave pulsations have been identified on timescales in the range 7-35 s \citep{collier_2023}. Imaging observations from HXR and microwaves have enabled us to localise the source of QPPs in this flare, thanks to the enhanced sensitivity of STIX onboard Solar Orbiter due to its close distance to the Sun (0.33 AU during this observation window) and complementary observations from EOVSA. As a result, many HXR sources have been resolved and have been shown to evolve in time. During the first phase of bursty emission, the HXR emission originates from multiple sources along the UV flare ribbons. The optically thin microwave sources typically appear at the location of a loop connecting HXR bright points to the northern ribbon. Spatially resolved microwave imaging analysis shows that pulsations originate from locations along the flare arcade with the clearest pulsations originating from a smaller region at the centre of the flare arcade (green box in Fig. \ref{fig:microwave_fitting}). These results are important as they tell us that the mechanism responsible for the observed QPPs must involve a 3D structure, i.e. the observations are not consistent with a stable loop with two fixed footpoints which oscillates due some perturbation of local plasma parameters. The fact that there are multiple footpoint sources means that the arcade must be considered in terms of a 3D structure that is changing in time. In contrast to previous studies \cite[e.g.][]{2003ApJ...595L.103K, Grigis_2005, Yang_2009, Inglis_2012} which typically show HXR sources moving in one direction along the flare ribbons, we find that the footpoints do not demonstrate a motion in a given direction, similar to the results obtained by \cite{2016SoPh..291.3385K}. In this case we can therefore excludes QPP models which involve reconnection triggered by wave propagation along the polarity inversion line (PIL) e.g. the slow mode propagation model proposed by \cite{Nakariakov_2011}.  

The results of the spectral analysis presented here are important as they constrain potential drivers to those that can modulate the electron spectral index significantly. The soft-hard-soft relationship has been found to hold for pulsations and peaks of flares in previous studies. For example, \cite{Grigis_2004} studied the spectral evolution of 24 M-class flares observed by RHESSI. They found the soft-hard-soft relationship applies to nearly all of the events studied and additionally found a power-law relation between the photon spectral index and the photon flux normalisation factor. They hence concluded that there is an intrinsic relationship between the flux and energy distribution of electrons for any elementary acceleration event (i.e. each pulsation of the flare). Furthermore, they concluded that the soft-hard-soft behaviour supports the idea that HXR pulsations or bursts represent an individual acceleration event, as is the case here. For this flare, we can exclude QPP generation models which modulate the observed emission post injection but do not accelerate particles significantly. An example of which is a sausage mode oscillation in a flare loop which can modulate the observed microwave and HXR emission by periodically trapping particles due to magnetic field variations transverse to the magnetic flux tube. Magnetic field variations can also accelerate particles via betatron acceleration. However, in this scenario, the electron spectral index remains unchanged \citep{Bogachev&Somov2007}. Furthermore, in this ideal scenario the observed HXR and microwave fluxes of each pulsation would be expected to be out of phase, which is not the case here (see Fig. \ref{fig:overview_figure}). Additionally, a coronal loop requires a stable environment to host an oscillation which is not the case during flares, especially during an eruption.

\subsection{Observational Limitations}
Here, we briefly discuss some of the observational limitations of this work. The main constraint of the observations presented here and a typical one of current HXR studies, stems from the fact that STIX is an indirect Fourier imager, similar to RHESSI. It hence suffers from dynamic range issues, such that faint sources are difficult to resolve and typically one can confidently plot contours at the 30\% levels and above in a given non-thermal image, but plotting lower levels is dependent on calibration and counting statistics. This comes from the fact that STIX only has 30 subcollimators (STIX has 32 detectors, so this excludes the background and coarse flare locator subcollimators) and therefore samples just 30 Fourier components or visibilities. As a result, STIX only resolves the brightest sources of HXR emission and hence it is possible that there are faint sources below the noise level of HXR emission in each frame presented in Fig. \ref{fig:early_impulsive_Earth}. 
In particular, subcollimators 3-10 were used in the reconstruction of the HXR sources presented in Fig. \ref{fig:early_impulsive_Earth}. This corresponds to only 24 Fourier components. With only 24 Fourier components the number of sources that can be reliably reconstructed is restricted. Image enhancement techniques such as Clean and MEM\_GE help to highlight the brightest sources in these maps, however, care must be taken when choosing the parameters of the algorithms such as beam width in the case of the Clean algorithm, for example. In order to overcome this issue in the future, a direct X-ray imager is required, such as the previously proposed Focusing Optics X-ray Solar Imager (FOXSI) \citep{foxsi_2014ApJ...793L..32K, christe_2023} and the need for such in QPP studies is highlighted by \cite{inglis2023}. The dynamic range limitation also makes it difficult to track the evolution of a particular source in time as the images obtained need a long enough integration time for flux modulation and therefore cannot be reconstructed at a sufficiently high time cadence. In this work, the observed sources are compared to those obtained in microwave which also suffers from dynamic range issues as it is a radio interferometer. However EOVSA is more sensitive to fainter sources with a dynamic range of $\sim$ 20:1 \citep{Gary_2018} and the upcoming Frequency Agile Solar Radiotelescope (FASR) will improve upon this further \citep{2023BAAS...55c.123G}. Nonetheless, useful information regarding the spatial origin of the brightest HXR sources and thus those that contribute most towards the observed variation are reconstructed.

\subsection{Potential Models}
For the SOL2022-03-30T17:21:00 X1.3 GOES class flare which exhibits non-stationary QPPs, the detailed analysis presented here has provided several constraints on the possible mechanism producing the observed pulsations. These are as follows:
\begin{itemize}
    \item The driving QPP mechanism must be able to synchronously modulate the observed HXR and microwave emission, with minimal time delay. This means that the energetic population of electrons which have different emission mechanisms in the two energy ranges must be modulated.
    \item The driving QPP mechanism should produce HXR emission (electron precipitation) in multiple locations along the flare ribbons which changes in time.
    \item The driving QPP mechanism must be able to modulate the electron spectral index significantly such that it is anti-correlated with the HXR flux (soft-hard-soft evolution with each pulsation).
\end{itemize}

According to a review article by \cite{Kupriyanova2020quasi-periodic}, there are three main categories of proposed QPP drivers: 
\begin{enumerate}
    \item Models involving the direct modulation of emitting plasma by MHD oscillations.
    \item Models in which the efficiency of energy release and particle acceleration is modulated by MHD waves.
    \item Models involving spontaneous quasi-periodic energy release.
\end{enumerate} 

From the constraints listed above, some scenarios can be excluded for the flare studied here. In particular, category (1) is not feasible in this case as the direct modulation of the emitting plasma (e.g. from sausage mode oscillations) cannot modulate the electron spectral index to the modulation depths we observe. Furthermore, in the ideal case of sausage mode oscillations, which is the most applicable for the pulsation timescales observed here, the microwave and HXR emission would be expected to be out of phase and this is contrary to what is observed in this case. Additionally, in order to host a standing wave mode, a stable loop like structure is required and the fact that we see multiple HXR sources at multiple footpoints which change in time is strong evidence against a standing MHD wave mode driver.

Based on the observations presented it is not possible to entirely exclude category (2) as a possible driver of the observed pulsations, however, it seems much more likely that category (3) is the responsible one in this case because we know that the accelerated electrons produce non-thermal bremsstrahlung emission at multiple locations along the flare ribbons and these locations change in time. If the energy release were due to a propagating MHD wave, the location of the energy release site would be expected to be more ordered than in reality \citep{Nakariakov_2011, Inglis_2012}. Furthermore, the complexity of the HXR emission appears to correspond to the complexity of the magnetic field geometry at a given time. During the eruption many HXR sources are observed, whereas post eruption the HXR source is concentrated in a simple two-footpoint configuration, as shown in Fig. \ref{fig:phase3_imaging}. As a result, in what follows we focus our discussion on mechanisms which belong to category (3). 

Fig. \ref{fig:early_impulsive_Earth} shows that throughout the early impulsive phase, when pulsations on timescales of $\sim 7-10$ s are observed, the HXR emission originates from locations along the UV ribbons. If we consider the 
brightest regions of HXR emission along the ribbon as individual sources of emission we notice that although there is some evolution in the exact location of each source and the relative brightness of individual sources, the main central HXR sources are present in most frames. In these frames there appears to be repeated electron precipitation in nearby loop structures. Furthermore, spatially resolved microwave imaging analysis shows that the pulsations originate from locations along the flare arcade. One such mechanism that could explain this phenomenon is particle acceleration from contracting magnetic islands which form in the flare current sheet following reconnection. \cite{Guidoni_2016} simulated this process for an eruptive flare and showed that the acceleration gain could produce the observed power law behaviour of flare accelerated electron spectra and in addition \cite{Guidoni_2016} showed that that the observed HXR emission would be "bursty" due to the stochastic nature of island formation. 

Another mechanism which could result in this type of observation is oscillatory reconnection \citep{McLaughlin_2008, thurgood2017, Karampelas2023}, in which the presence of a nonlinear fast magnetoacoustic shocks in the vicinity of a magnetic X-point causes the system to evolve through a series of vertical and horizontal current sheets which leads to oscillatory reconnection. However, oscillatory reconnection is a relaxation process and is expected to show a damped signal \citep{McLaughlin_2012a}, which is at odds with the observations presented here. Furthermore, it is difficult to understand the feasibility of this mechanism in relation to the eruption and the 3D nature of the flare arcade.  

An aspect that is clear from this analysis is that the source of these pulsations in HXR and microwave emission, is a complex 3D structure. The observed pulsating sources extend along the UV ribbons and thus clearly originate from various acceleration sites. Due to the complexity of the arcade's magnetic structure, there could be multiple reconnection sites which extend along the length of the PIL. In this case, pulsation timescales would simply arise from the unique flare arcade configuration.  

\section{Conclusions} \label{sec:conclusions}
In this work the X1.3 GOES class flare, SOL2022-03-30T17:21:00 has been analysed. This flare displays non-stationary QPPs in the HXR and microwave emission observed with STIX and EOVSA. The pulsations are on timescales evolving from $\sim 7$ s in the impulsive phase to $\sim 35 $s in the later flare stages, after the SXR peak. Detailed analysis of the HXR and microwave source locations revealed that the source of pulsations is changing in time. Multiple HXR sources are present along the UV flare ribbons throughout the impulsive phase. It is found that the electron spectral index inferred from HXR observations is anti-correlated with the flux observed in HXR and microwave and obeys a soft-hard-soft evolution with each sub-peak. This indicates that the pulsations are related to the periodic injection and/or acceleration of electrons. As a result, models involving spontaneous quasi-periodic energy release have been identified as the driver of the observed pulsations, in particular those involving multiple electron acceleration sites along the flare arcade. This work demonstrates that coordinated microwave and HXR observations of solar flares enable us to probe the feasibility of proposed models of time variability in flare emission. As such, future coordinated observations between HXR imagers including STIX, ASO-S/HXI \citep{HXI_Zhang2019}, Aditya-L1/HELIOS \citep{2017aditya} and EOVSA will advance our understanding of rapid variation in HXR and microwave emission during solar flares greatly.

\begin{acknowledgements}
           Solar Orbiter is a space mission of international collaboration between ESA and NASA, operated by ESA. The STIX instrument is an international collaboration between Switzerland, Poland, France, Czech Republic, Germany, Austria, Ireland, and Italy. The Expanded Owens Valley Solar Array (EOVSA) is a ground based facility operated by the New Jersey Institute of Technology (NJIT), under the support of NSF grant AGS-2130832 and NASA grant 80NSSC20K0026 to NJIT.
            
            HC, AFB and SK are supported by the Swiss National Science Foundation Grant 200021L\_189180 for STIX.
            L.A.H is supported by an ESA Research Fellowship. S.Y. is supported by NASA grant 80NSSC20K1318 and NSF grant AST-2108853 awarded to NJIT.
            WA is supported by NASA grant 80NSSC20K1283.
            VP acknowledges support from NASA ROSES Heliophysics Guest Investigator program grant 80NSSC20K0716 and  NNG09FA40C ({\it IRIS}).
            This work was additionally supported by the ESA Archival Research Visitor Programme.
            We would like to thank Brian R. Dennis and Paolo Massa for informative discussions which have significantly enhanced this work.

\end{acknowledgements}

% WARNING
%-------------------------------------------------------------------
% Please note that we have included the references to the file aa.dem in
% order to compile it, but we ask you to:
%
% - use BibTeX with the regular commands:
%   \bibliographystyle{aa} % style aa.bst
%   \bibliography{Yourfile} % your references Yourfile.bib
%
% - join the .bib files when you upload your source files
%-------------------------------------------------------------------

\bibliographystyle{aa} % style aa.bst
\bibliography{XXX.bib} % your references Yourfile.bib

\begin{appendix}
\section{Reliability of Hard X-Ray Imaging} \label{apdx:vis_fits}
\begin{figure*}
    \includegraphics[width=\textwidth]{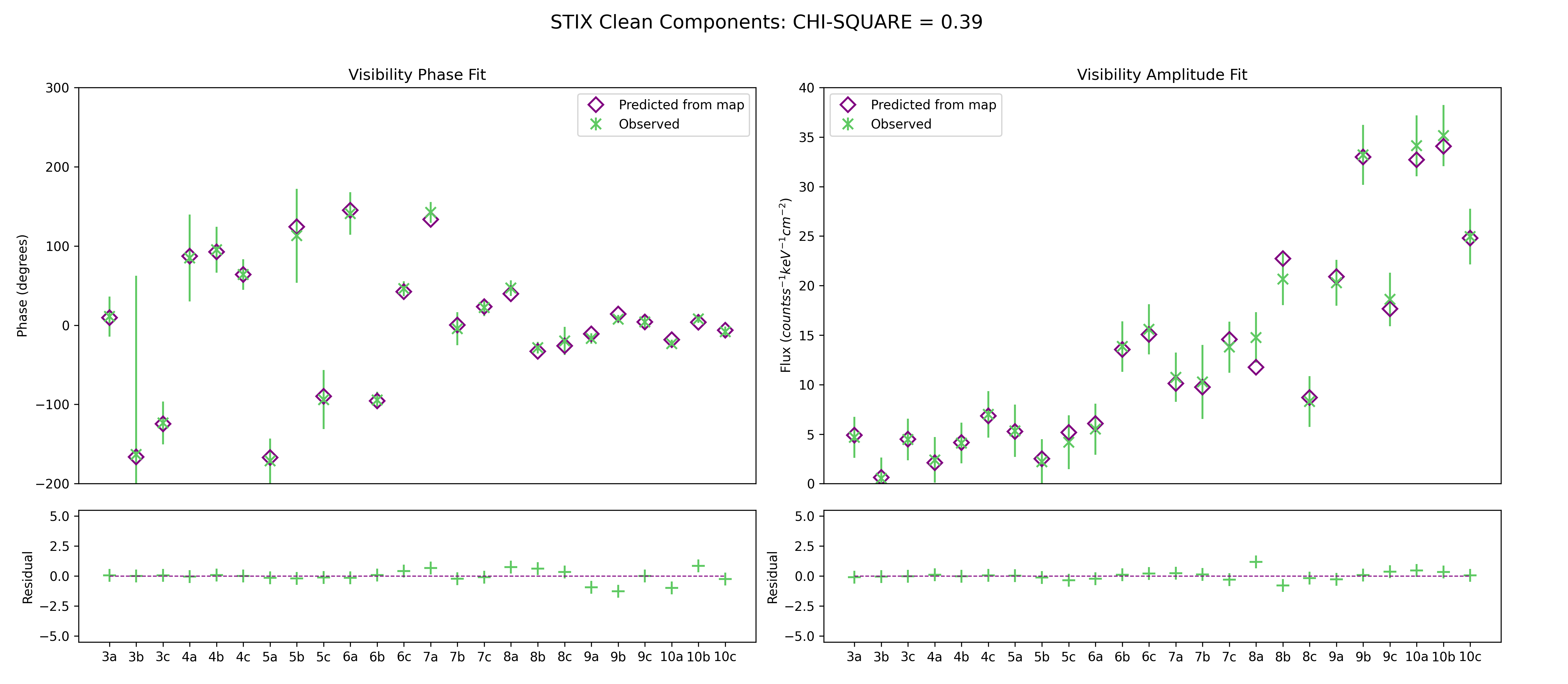}
    \caption{Figure shows the predicted visibilities from the reconstructed Clean map shown in Fig. \ref{fig:early_impulsive_solo} compared to the measured visibilities.}
    \label{fig:clean_vis_fit}
\end{figure*}

\begin{figure*}
    \includegraphics[width=\textwidth]{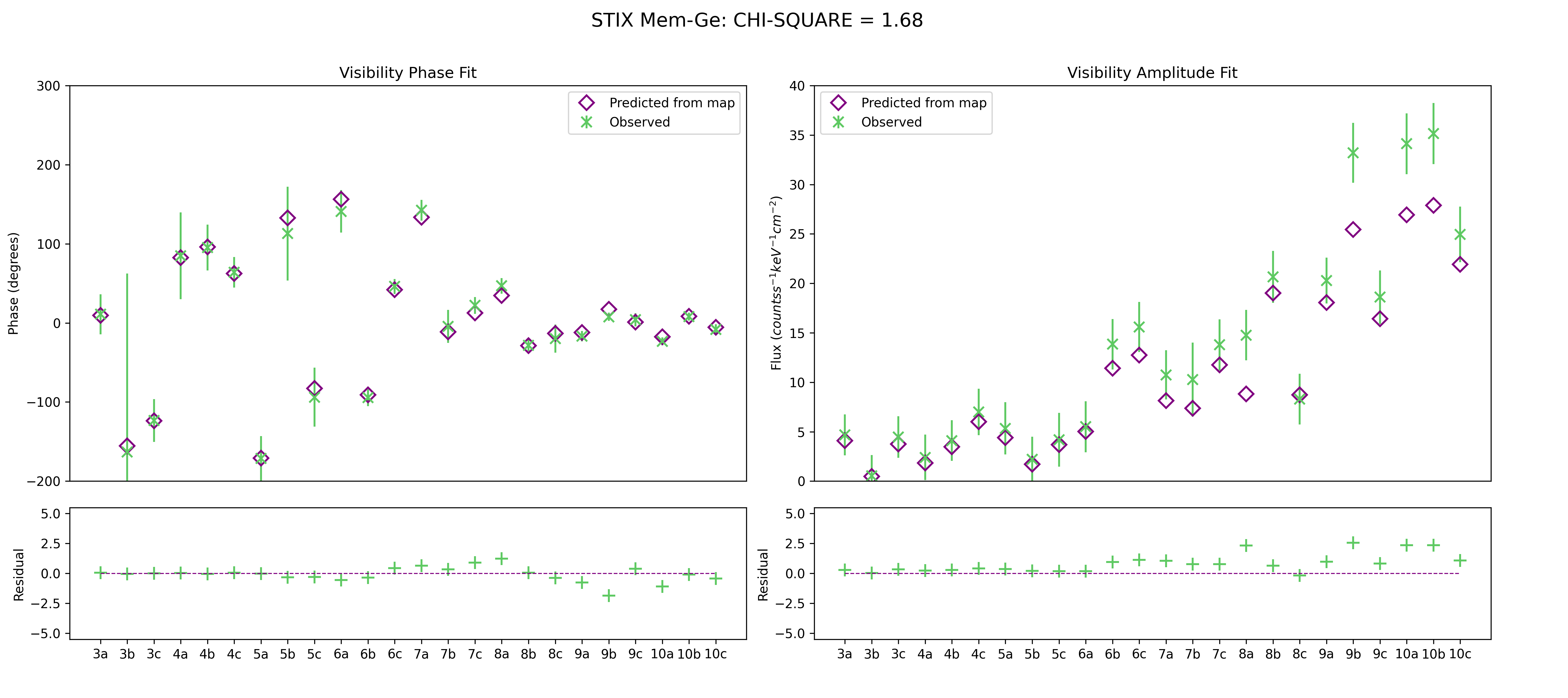}
    \caption{Figure shows the predicted visibilities from the reconstructed MEM\_GE map shown in Fig. \ref{fig:early_impulsive_solo} compared to the measured visibilities.}
    \label{fig:mem_vis_fit}
\end{figure*}

Fig. \ref{fig:clean_vis_fit} and Fig. \ref{fig:mem_vis_fit} show the visibility phase and amplitudes for the observed maps and those predicted from the reconstructed maps for both the Clean and MEM\_GE non-thermal images shown in Fig. \ref{fig:early_impulsive_solo}. Both demonstrate a good fit to the data; Clean obtains a reduced Chi-square $\chi_{\nu}^2 = 0.39$ and MEM\_GE achieves $\chi_{\nu}^2 = 1.68$. Additionally, it is notable that the visibility amplitude of subcollimators 9b, 10a \& 10b are poorly fit by MEM\_GE.

\section{Spectral Fitting}    \label{apdx:example_spectral_fit}
\begin{figure*}
    \includegraphics[width=\textwidth]{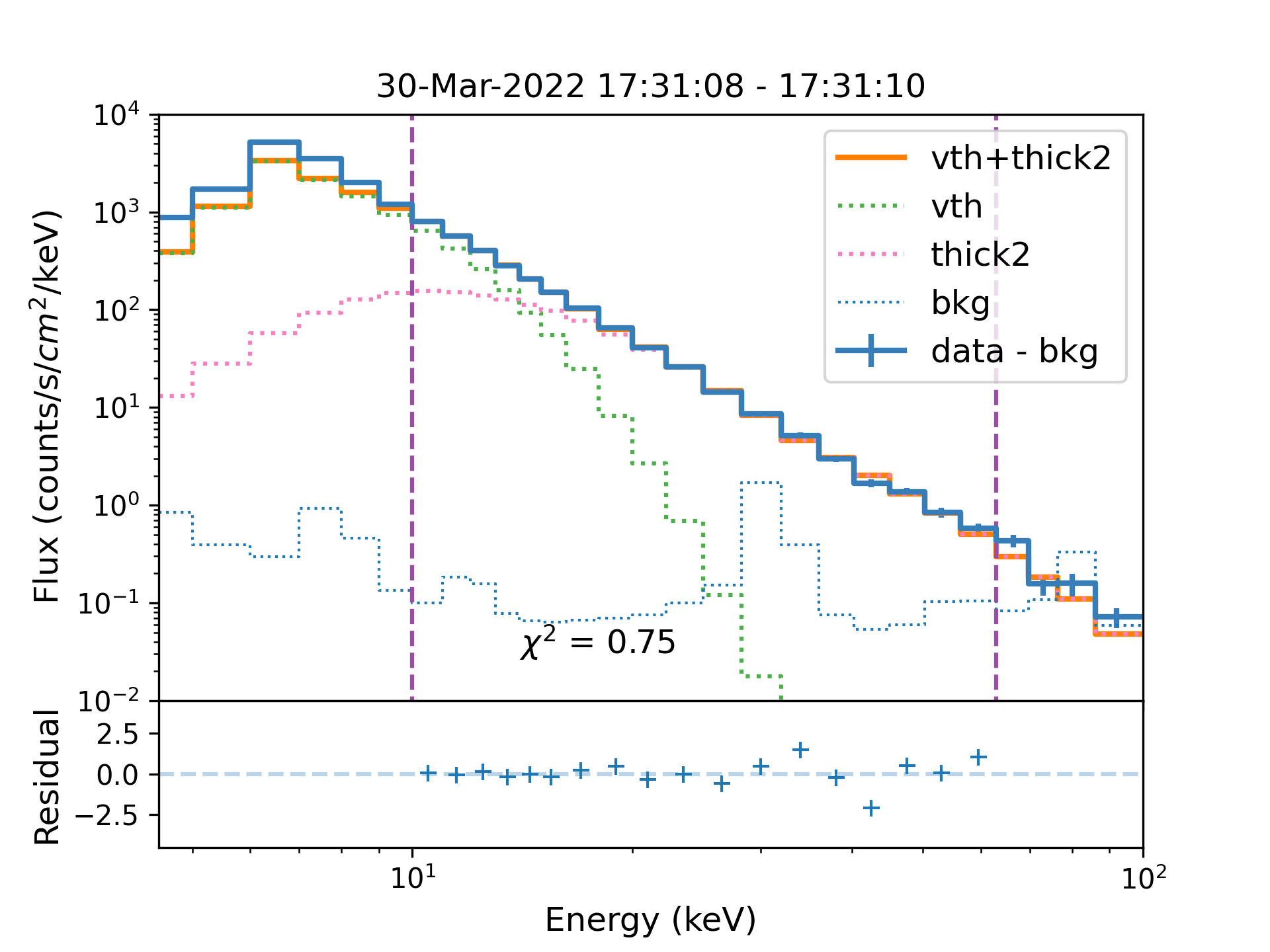}
    \caption{Example spectral fit involving a thermal and non-thermal thick target bremsstrahlung component (\texttt{f\_vth.pro} and \texttt{f\_thick2.pro}) for the time interval 17:31:08-17:31:10. The fitting interval is indicated by the two dashed lines at 10 and 63 keV. The fit gives a reduced chi-square value of $\chi_{\nu}^2 = 0.75$. }
    \label{fig:example_spectral_fit}
\end{figure*}

Fig. \ref{fig:example_spectral_fit} shows a an example spectral fit to the measured data for the time interval 17:31:08-17:31:10. A thermal and non-thermal thick target bremsstrahlung model is fit.

\end{appendix}

\end{document}